# NON-NUCLEON METASTABLE EXCITATIONS IN NUCLEAR MATTER AND $e^-$ CATALYSIS AS A QUARK-CUMULATIVE MECHANISM FOR INITIATING LOW-ENERGY NUCLEAR CHEMICAL PROCESSES: PHENOMENOLOGY

S. F. Timashev

National Research Nuclear University, Moscow Engineering Physics Institute (MEPhI), Moscow, Russia

e-mail: serget@mail.ru

The present study demonstrates that the mechanism of initiation of low-energy nuclear chemical processes under conditions of low-temperature non-equilibrium deuterium and protium-containing glow discharge plasma is similar to the previously studied cumulative mechanism of initiation of nuclear processes in the collision of relativistic particles (protons) with target atomic nuclei. This process results in the formation of high-energy products that significantly exceed the kinematically resolved region in the pulse space for two-particle collisions "nucleus–target's nucleus." The cumulative effect in this case is associated with the initiation of non-nucleonic metastable excitations in nuclear matter during relativistic collisions leading to the formation of a group of quarks from different nucleons within the nucleus. In low-energy nuclear chemical processes, the initiation of quark-cumulative processes in nuclear matter occurs through interaction of nuclei with electrons with high kinetic energies on a chemical scale, typically $E_e \sim 3\text{–}5$ eV. Experiments and available literature data suggest that the metastable excitations of nuclear matter containing three "free" quarks which occur during such collisions are associated with quark-cumulative effects, leading to the radioactive $\alpha$- and $\beta$- decay of elements. This phenomenon is observed during laser ablation of metals in aqueous media containing radioactive elements and in the artificial radioactivity of initially non-radioactive isotopes in cathodes exposed to low-temperature non-equilibrium deuterium- and protium-containing plasma flows during glow discharge.

INTRODUCTION

The first evidence of non-nucleon degrees of freedom in atomic nuclei, now associated with quarks, emerged in 1957. In Dubna, using a proton beam with an energy of 660 MeV, G.A. Leksin and coworkers observed an unexpectedly high yield of protons in the rear hemisphere during the study of elastic scattering of protons on deuterons with large momentum transfers. Simultaneously, the group of M.G. Meshcheryakov observed an unusually high yield of deuterons in the "forward" direction when irradiating light nuclei with protons of energy of 675 MeV [1–3]. These observations led to the suggestion [1] that "the emission of fast fragments from nuclei can be considered as a result of quasi-elastic interactions of an incident nucleon with a group of nucleons in the nucleus that is strongly bound at the moment of collision." To explain the observed effects, in the same year (1957) D.I. Blokhintsev [4] proposed a hypothesis about the existence of density fluctuations in nuclear matter within the nucleus, later termed "fluctons." In further development of these concepts, Baldin [5] proposed a hypothesis regarding cumulative nature of the formation of density fluctuations in nuclear matter. According to this hypothesis, particles generated during the collision of a relativistic nucleus with a target can receive energy far exceeding the energy per nucleon of the incident nucleus, contrary to the conservation laws of energy and momentum in free particle-particle collisions. This phenomenon occurs when an incident relativistic particle (such as a proton, in the examples considered) interacts effectively with more than one nucleon of the target nucleus. This effect seems to be a consequence of non-nucleon, "parton-quark" interactions, occurring at distances within the nucleus that are much larger than the size of a nucleon. Therefore, as suggested [5], the study of such a quark-cumulative effect can provide insights into the interaction of quarks at large distances, which is crucial for understanding "quark confinement." It should be noted that subsequent experimental studies of the cumulative effect conducted by the of V.S. Stavinsky group [3] in Dubna provided support for Baldin's hypothesis. Hence, at present, the quark-cumulative Baldin effect in nuclear matter during collisions of relativistic nuclei with a target is now interpreted as a consequence of "non-two-particle" collision mechanism, involving the initiation of non-nucleon metastable excitations associated with the "sharing" of quarks among a group of nucleons. Due to the nature of these collisions, anomalously high-energy products are generated, the energy of which can significantly exceed the kinematically allowed region in momentum space for two-particle "nucleus–target's nucleus" collisions.

However, recent research findings in the field of Low-Energy Nuclear Reactions (LENR) [6–11] or "Nuclear Chemical processes" [12–17] challenge and broaden the traditional understanding of cumulative phenomena in nuclear matter as a phenomenon associated only with the relativistic energies of colliding nuclei. These processes, spanning a wide range of nuclear phenomena such as *α*- and *β*- radioactive decays, nuclear fusion, transmutation of nuclei, and initiation of artificial radioactivity have been investigated extensively over the past decade. Indeed, the initiation energy of these processes was found to be several orders of magnitude lower than the typical energies (in the range of MeV) required for the initial particles (protons, α-particles, and other nuclei) to overcome Coulomb energy barriers and trigger nuclear processes during the interaction of falling particles with the atomic nuclei of the targets under study. Furthermore, "low-energy" nuclear reactions are notable for their minimal production of ionizing radiation and neutron emission, which are characteristic features of conventional nuclear transformations requiring specialized safety precautions. Both of these factors have led to cautious and, at times, skeptical attitudes among some specialists towards the results of such studies, despite the publication of relevant articles in reputable specialized journals in recent years (for example, [8, 9, 16]).

Possible mechanisms for the implementation of nuclear chemical processes were proposed, along with an explanation of the physical reasons for the manifestation of these anomalies in a fairly general case of initiation of such processes, namely, under conditions of low-temperature nonequilibrium plasma formed during laser ablation of metals in aqueous media [12-15] and during a glow discharge in protium and deuterium containing gaseous media [16, 17]. First of all, as it turned out, the ideas already introduced earlier into nuclear physics by A.M. Baldin [2, 5] were introduced about the possibility of the formation of non-nucleon metastable excitations in nuclear matter initiated by interaction with atoms (ions) and electrons of low-temperature plasma. Unlike relativistic particles (protons) in Baldin's work, the interactions involve electrons with high, albeit chemically scaled, kinetic energies $E_e \sim 3-5$ eV. These exciting effects on the electronic subsystems of atoms (ions) determine the interaction of the inner-shell electrons of the corresponding atoms
(ions) with atomic nuclei.

It was assumed that if the original nucleus $_Z^A N$ (characterized by $Z$ and $A$, atomic number and mass number of the nucleus N, respectively) does not undergo $K$-capture, then the interaction of the initiated electron with the near-surface region of the nucleus leads to the emission of neutrinos and the formation of a vector $W^-$ boson. When interacting with $u$-quark from one of the protons of nuclear matter, a $d$-quark is formed. However, due to the lack of mass for such an isotope, a neutron is not formed in nuclear matter. Therefore, the nucleon structure of the nucleus M, formed during this process:

$$_Z^A N + e_{he}^- \rightarrow {}_{Z-1}^A M_{isu} + \nu, \tag{1}$$

is locally disrupted (resulting in three "free" quarks!), and the nucleus M enters a metastable state of "inner shake-up" or *isu*-state. The subscript when representing an electron on the left side of Eq. (1) indicates (high energy) the activated nature of this stage of the process, and the subscript in the notation of the nucleus on the right side of Eq. (1) indicates its metastable state.

The subsequent relaxation decay of this nucleus (we will define these nuclei as "*isu*-nucleus"):

$$_{Z-1}^{A}M_{isu} \rightarrow {_{Z}^{A}N} + e^- + \tilde{\nu}, \tag{2}$$

like reaction (1) is determined by weak nuclear interactions. For this reason, the general process of the considered nuclear initiation and subsequent β-nuclear decay in the *isu*-state involves the inelastic scattering of an electron on a nucleus through the weak interaction channel with the emission of a neutrino-antineutrino pair. The energy threshold for this process is 0.3 eV [18]. The latter determines the indicated values of kinetic energy of electrons $E_e$ in a nonequilibrium low-temperature plasma, sufficient for the implementation of nuclear chemical transformations considered in previous works [12–17]. When the kinetic energy of electrons increases above the specified values, the probability of initiation of the nuclear chemical processes under consideration decreases due to the predominant loss of electron energy in the plasma to the ionization processes of atoms and ions.

The introduction of ideas about the existence of non-nucleon metastable excitations in nuclear matter and the possibility of implementing quark-cumulative mechanisms for initiating nuclear processes is the basis for understanding the physical essence of nuclear chemical transformations in the conditions of a protium-or deuterium-containing glow discharge plasma. In this case, a fairly soft controlled mechanism of the impact on the state of nuclear matter and nuclear processes is implemented, which can be defined as "$e^-$ – catalysis." Here, we will consider two options for this type of catalytic transformation. First type of the $e^-$ – catalys is consists in the direct initiation of quark-cumulation processes in nuclear matter of initially α- and β-radioactive nuclei, whose charge decreases by one after the formation of a trio of "free" quarks. In the second type of the $e^-$ – catalysis, the simplest *isu*-nuclei, namely *isu*-neutron $^1n_{isu}$ and *isu*-dineutron $^2n_{isu}$, which are formed, according to Eq. (1), during the interaction of high-energy electrons with protons $^1_1H$ or deuterons $^2_1H$, respectively, play an active role. Since these neutral nuclei have no barriers to interacting with other nuclei, the question arises about the characteristic times of existence of these *isu*-states to understand how purposefully the nuclei $^1n_{isu}$ and $^2n_{isu}$ formed in low-temperature plasma can be utilized in nuclear physics.

To resolve this issue, a special experiment was previously carried out, the results of which [15] made it possible to understand how nucleus *isu*-state $^2n_{isu}$ may manifest itself in the process of tritium synthesis $t^+$ during laser ablation of metals in heavy water, when laser exposure resulted in the formation of a low-temperature non-equilibrium plasma in the area adjacent to the metal plate. It was believed that the process of tritium synthesis should have occurred through the interaction of a tritium nucleus $t^+$ with nucleus $^2n_{isu}$:

$$d^+ + {}^2n_{isu} \to t^+ + n + Q(3.25\text{MeV}),\tag{3}$$

where $n$ is the neutron. In this case, along with Eq. (3), the following process should also occur:

$$d^+ + {}^2n_{isu} \to {}^3_2He + n + e^- + \tilde{v} + Q(3.27\text{MeV}),\tag{4}$$

which is determined, like process (3), by weak nuclear interaction.

The possibility of formation of hypothetical *isu*-trineutron ${}^3n_{isu}$ during the reaction of electrons with tritium $t^+$ nuclei was also postulated [15]:

$$t^+ + e^-_{he} \to {}^3n_{isu} + v.\tag{5}$$

Rest mass of the introduced neutral nucleus ${}^3n_{isu}$ was assumed to be equal to the rest mass of the tritium atom. It is hypothesized that the formation of a nucleus ${}^3n_{isu}$ should be involved in the process of initiated decay of tritium nuclei under conditions of laser ablation of metals in aqueous media found [15], along with the process of synthesis of tritium nuclei:

$$t^+ + e^-_{he} \to {}^3n_{isu} + v \to {}^3_2He + 2e^- + v + 2\tilde{v} + Q(0.019\text{MeV}).\tag{6}$$

According to the data presented [15], the activity of tritium in heavy water exceeded the background activity of the original system by an order of magnitude and by three orders of magnitude when a cathodic bias was applied to the metals used. Moreover, the entire set of results obtained could be understood by assuming that the half-lives $T_{1/2}$ of both nuclei ${}^2n_{isu}$ and nuclei ${}^3n_{isu}$ are at least 10 min. The same half-lives are characteristic of nuclei ${}^1n_{isu}$, as may follow from subsequent experiments [16]. It should be noted here that the given value of the half-life of nucleus ${}^3n_{isu}$ was found to be many orders of magnitude less than the half-life $T_{1/2}$ of the tritium nucleus ($T_{1/2}$ = 12.3 years), which indicates the initiating nature of tritium decay under conditions of low-temperature non-equilibrium plasma.

ELECTRON FACTOR IN THE INITIATION OF THE DECAY OF INITIALLY RADIOACTIVE ISOTOPES: 1st TYPE $e^-$ CATALYSIS

The process of decay of radioactive isotopes was examined more comprehensively in previous studies [12, 14, 19]. It was suggested that the formation of metastable *isu*-states in nuclear matter during process (1) led to a violation of the general stability of the nucleus ${}^A_{Z-1}M_{isu}$, which is determined by the radial component of the pressure tensor. This tensor is formed by zero-point oscillations of the electromagnetic (EM) vacuum [20] and is associated with changes in the boundary conditions for the components of the electric field strength vector of the EM vacuum on the surface of this nucleus. As an indicator of the instability of the resulting nucleus $_{Z-1}^{A}M_{isu}$, which determines the subsequent rate of its radioactive decay with the emission of daughter products, the absolute value of the structural energy deficit $\Delta Q_{NM}^{A}$ ($\Delta Q_{NM}^{A} < 0$) of this nucleus was considered, which is missing to form the basic state of nuclear matter in the nucleus $_{Z-1}^{A}M_{isu}$; therefore, $\Delta Q_{NM}^{A} = (m_{_{Z-1}^{A}M_{isu}} - m_{_{Z-1}^{A}M})c^2$. At the same time, for the mass of the nucleus $_{Z-1}^{A}M_{isu}$ we accepted $m_{_{Z-1}^{A}M_{isu}} = m_{_{Z}^{A}N} + m_e$, where $m_{_{Z}^{A}N}$ is the mass of the nucleus $_{Z}^{A}N$ and $m_e$ is the rest mass of the electron.

The aforementioned conclusion was drawn based on the findings of study [12], which investigated the initiation of radioactive decay of the isotope $_{92}^{238}U$ during laser ablation of metal samples of different nature in an aqueous solution of uranyl chloride. This process creates a low-temperature non-equilibrium plasma in a vapor environment adjacent to the metal surface. It was hypothesized that the effects of high-energy chemical electrons ($E_e$ ~ 3–5 eV [21]) from such plasma on the electronic subsystem of atoms $_{92}^{238}U$ lead to oscillations of the electron subsystem. According to [20], this subsystem is regarded as a unified system of "entangled" electrons formed during the Casimir polarization of the EM vacuum by electrons. These forced oscillations of the single electronic subsystem of the atom $_{92}^{238}U$ initiate the interaction of this subsystem with the atomic nucleus, resulting in process (1) with the formation of an unstable nucleus known as "isu-protoactinia." Subsequently, $\beta^-$-nuclear decay of nuclei $_{91}^{238}Pa_{isu}$ occurs, resulting in the formation of thorium-234 and helium-4 nuclei as decay products of the original uranium-238 nucleus:

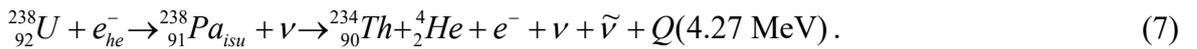

$$_{92}^{238}U + e_{he}^- \rightarrow {}_{91}^{238}Pa_{isu} + \nu \rightarrow {}_{90}^{234}Th + {}_{2}^{4}He + e^- + \nu + \tilde{\nu} + Q(4.27 \text{ MeV}). \quad (7)$$

In this case, the effective constant $k$ of the rates of such initiated nuclear decays $_{92}^{238}U$ related to the half-life of the nucleus $T_{1/2} = \ln 2 / k$, as shown in [12], increased by 9 orders of magnitude, which indicated the implementation of the $e^-$-catalysis. Shortage of the structural energy $\Delta Q_{UPa}^{238}$ of the forming *isu*-protoactinia nucleus constitutes $\Delta Q_{UPa}^{238} \approx -3.46$ MeV. At the same time, the constant $k$ of the rate of the nucleus $_{92}^{235}U$ decay (in this case $\Delta Q_{UPa}^{235} \approx -1.41$ MeV) was practically not changed within the error limits.

An unexpected result was obtained during experiments conducted with beryllium and gold samples. It was found that beryllium nanoparticles formed in solution after completion of laser ablation for an hour exhibited an abnormally high rate of formation of thorium-234 nuclei for more than 500 days after completion of laser ablation. For gold nanoparticles, the same effect was quantitatively less pronounced. Half-life for nuclei $^{238}_{91}Pa_{isu}$ initiated under laser ablation conditions producing thorium-234 was 2.5 years. This phenomenon could naturally be associated with the accumulation of *isu* protoactinia nuclei in beryllium and gold nanoparticles during laser ablation, which lasted only an hour.

The introduced parameter $\Delta Q^A_{NM}$ should be considered as an indicator of effectiveness of the $e^-$-catalysis in initiation not only α-decay but also β-decay of radioactive nuclei with the formation of final nuclei in a certain excited or ground state. This follows, in particular, from the findings of study [22], where data [23] on the study of $\beta^-$-active nuclei $^{60}_{27}Co$, $^{137}_{55}Cs$ and $^{140}_{56}Ba$ were analyzed under conditions of their initiation in a concentrated mass of various types of metabolically active microorganisms. Processes of the $e^-$-catalysis for $\beta^-$-decays of these nuclei were represented as follows [22]:

$$^{60}_{27}Co + e^-_{he} \to {}^{60}_{26}Fe_{isu} + v \to {}^{60}_{28}Ni + 2e^- + v + 2\tilde{v} + Q(2.82 \text{ MeV}), \tag{8}$$

$$^{137}_{55}Cs + e^-_{he} \to {}^{137}_{54}Xe_{isu} + v \to {}^{137}_{56}Ba + 2e^- + v + 2\tilde{v} + Q(1.18 \text{ MeV}), \tag{9}$$

$$^{140}_{56}Ba + e^-_{he} \to {}^{140}_{55}Cs_{isu} + v \to {}^{140}_{57}La + 2e^- + v + 2\tilde{v} + Q(1.05 \text{ MeV}). \tag{10}$$

In these cases, the magnitude of the structural energy deficit $\Delta Q^A_{NM}$ (in the future, for simplicity, we will omit the upper and lower indices of this value if they are obvious from the context), which the nuclei $^{60}_{26}Fe_{isu}$, $^{137}_{54}Xe_{isu}$ and $^{140}_{55}Cs_{isu}$ lack in the "inner-shake-up" state before the formation of the basic state of nuclear matter characteristic of daughter nuclei $^{60}_{26}Fe$, $^{137}_{54}Xe$ and $^{140}_{55}Cs$, are –0.237, –4.17, and –6.22 MeV, respectively. It can be expected that the most pronounced effect of electrons on the $\beta^-$-decay of nuclei in low-temperature plasma will occur when the forming nuclei in the *isu*-state exhibit the greatest "mismatch" in the absolute value of the deficit $\Delta Q$ structural energy. Therefore, in the cases under consideration, the effect of accelerating radioactive decay should be most pronounced for nuclei $^{137}_{55}Cs$ and $^{140}_{56}Ba$, whereas for nuclei $^{60}_{27}Co$ it should be minimal. The experimental data presented [23] regarding the initiated decays of the studied $\beta^-$-active nuclei $^{137}_{55}Cs$, $^{140}_{56}Ba$ and $^{60}_{27}Co$ are quite consistent with this conclusion: the half-lives $T_{\frac{1}{2}}$ of nuclei $^{137}_{55}Cs$ and $^{140}_{56}Ba$ equal to 30.1 years and 12.8 days, respectively, decreased to 380 days and 2.7 days, while the half-lives of $^{60}_{27}Co$ equal to 1925 days remained virtually unchanged.

In this scenario, a natural question arises: if the energy parameter $\Delta Q^A_{NM}$ for the initial radioactive nucleus is positive, as is the case with electron or *K*-capture, to what extent the activation of such nuclei upon collision with an electron and the formation of the *isu*-state of the nuclear structure in such nuclei is also a necessary condition for radioactive decay? A positive answer to this question follows from the findings of work [24], in which a 50-year-old mystery was solved. The physical reasons for the anomalies in the apparent decrease in the fundamental coupling constant when calculating the probabilities of processes of the *β*-decay, including electron capture, were determined. The data are relevant if we focus on the predictions of theoretical calculations based on the analysis of experimental data on *β*-decay of free neutrons. The anomalies arose from the fact that the *β*-decays of unstable atoms occurred approximately 25% less frequently than was expected from the above theoretical calculations. In [24], model calculations on powerful supercomputers using quantum chemistry methods and taking into account intranuclear interactions of two, three or more nucleons demonstrated that in processes realized through weak nuclear interaction, the *β*-decay and electron capture simultaneously involve two nucleons, leading to the emergence of strong long-range correlations in nuclear matter. Essentially, this entails the implementation of the quark-cumulative Baldin effect in nuclear matter, not only in the *β*-decays but also in electron capture processes, when $\Delta Q_{NM}^{A} > 0$. Therefore, the factor determining the dynamics of relaxation of the *isu*-state of nuclear matter, is the parameter $\left|\Delta Q_{NM}^{A}\right|$ for *α*- and *β*-decays, when the relaxation dynamics, in accordance with the principle of least action, is oriented towards the formation of daughter nuclei as decay products of the original nucleus, and a positive value $\Delta Q_{NM}^{A}$ for electron capture, when the relaxation dynamics of nuclear matter is oriented towards completing the formation of the nucleon structure of the daughter nucleus. Nature, according to Aristotle, indeed, "does nothing in vain and in all her manifestations chooses the shortest or easiest path." In accordance with this principle, radioactive decays are realized with sufficiently "deep" excitation of nuclear matter, when quarks of different nucleons participate in its relaxation rearrangements, thereby fully manifesting the phase volume of nuclear matter and materializing the principle of least action.

In [13, 14] the Feynman diagrams (see Fig. 1) of initiated $\beta^{-}$ - and $\beta^{+}$ -decays, *K*-capture, as well as initiated *α*-decays, illustrating the dynamics of initiation of the corresponding decays, are discussed. It should be borne in mind that these processes are realized on spatial scales exceeding the scales of individual nucleons and, possibly, covering a significant part of the volume of the nucleus, specifically quarks of different nucleons. This type of spatial dynamics corresponds to the implementation of Baldin's general idea about the quark-cumulative mechanism of energy concentration on individual degrees of freedom in nuclear matter.

In light of evolving concepts regarding the initiation of the $\beta^-$- and $\beta^+$-decays, as well as $\alpha$-decay under conditions of the $e^-$-catalysis, a natural question arises: to what extent will the participation of initiating electrons in the processes of radioactive decay change the well-established ideas about the shape of the energy spectra of fixed particles – the continuous spectrum of electrons at the $\beta$-decay and discrete spectrum of α-particles at the α-decay? Since in the case of an initiated -decay the entire energy of the process is distributed between two electrons, a neutrino and two antineutrinos (with the contribution of the recoil nuclei energy being negligible), and the energy of only one electron is experimentally recorded, the study effectively investigates the spectrum of the inclusive process, namely, the energy spectrum of only one of the final particles. Consequently, the remaining set of particles is effectively treated as a single undetectable particle in the experiment. Therefore, the energy spectrum of the detected electron exhibits the typical shape characteristic of the spectrum of the $\beta$-decay type. We believe that due to the inclusive nature of the process, in none of the studied $\beta$-decays, to the best of the author's knowledge, an $\beta$-particle with kinetic energy equal to the decay energy has been recorded, in contrast to the energy spectrum of $\alpha$-particles recorded at the $\alpha$-decay, which has the form of a peak with a maximum at the energy of the process under study. The involvement of the initiating electron in this process may manifest as a contribution to the usually observed peak broadening and an increase in the initial electron background. Clearly, to establish the adequacy of the developed ideas about the activated nature of the $\beta$-decay processes, an analysis of these processes involving the recording of both emitted electrons and the realized delay times between their successive emissions is necessary.

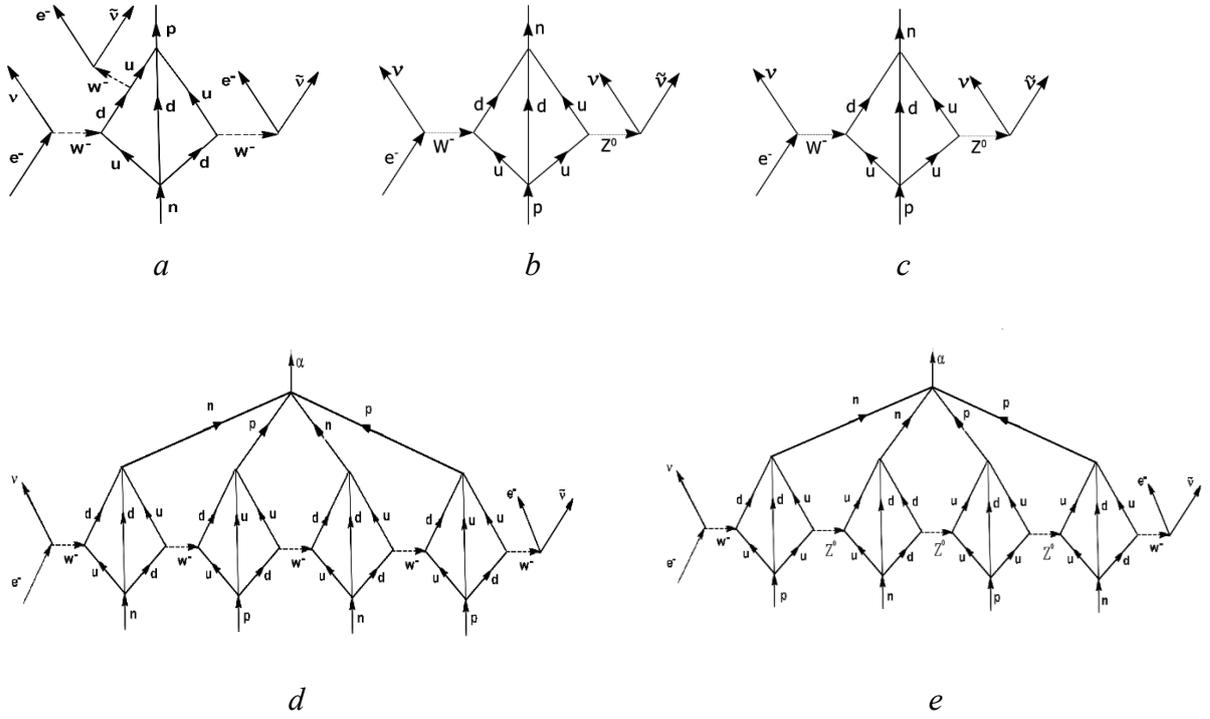

**Fig. 1.** Feynman diagrams of the triggered (a) $\beta^-$-decay, (b) $\beta^+$-decay, (c) K-capture, and (d, e) initiated $\alpha$-decay.

When analyzing processes presented on Fig. 1, it must be borne in mind that weak nuclear interactions are not as weak as is often believed. The value of the corresponding dimensionless constant $\alpha_F$ almost an order of magnitude greater than $\alpha_e$ = 1/137 of the fine structure constant [20]. Indeed, this already follows from $a_Z = 2^{1/2}\hbar/m_Z c \approx 3.3 \ 10^{-16}$ cm, which is characteristic size associated with the mass of the intermediate $Z^0$ vector boson ($m_Z$ = 91.2 GeV/c² = 1.62 ×10$^{-22}$ g), and $G_F = 1.17 \cdot 10^{-5}(\hbar c)^3/(\text{GeV})^2$, which is the Fermi constants of four-fermion interaction. In this case, the square of the "elementary charge of the weak nuclear interaction" is equal to $q_F^2 \equiv G_F/a_Z^2$; therefore, for the dimensionless value of the constant $\alpha_F$ of weak interaction, we have: $\alpha_F = \dfrac{q_F^2}{\hbar c}$ $\approx 4.9 \ 10^{-2}$, and $\alpha_F/\alpha_e$ = 6.7. Unfortunately, in the literature, the dimensionless constant of weak nuclear interaction is often estimated using the proton mass as a normalizing factor, following the tradition that persisted until the 1980s when intermediate vector bosons were discovered at CERN in 1983. However, the mass of the $Z^0$ vector boson is almost 100 times greater than the proton mass. As a result, the value of the constant $\alpha_F$ is underestimated by almost four orders of magnitude. This oversight significantly complicates the understanding of the scale of universality in the manifestations of weak nuclear interactions.

In connection with the given estimate of the dimensionless constant $\alpha_F$, one more clarification should be made related to the use of the four-fermion interaction constant introduced by Fermi in 1933 for this estimate $G_F$, and not the quantities corresponding to the three-particle vertex parts involving intermediate vector bosons discovered half a century later. At the same time, to introduce the characteristic size at which processes associated with weak nuclear interactions take place, which is necessary to determine the "charge" of the weak interaction, the mass of the intermediate $Z^0$ vector boson is used. According to the author, in discussing the $\beta$-decay process, the need to use such eclecticism indicates the key point of the process, which consists in the fact that during the interaction between the vector boson $W^-$ with the $u$-quark of one of the protons of nuclear matter simultaneously with the formation of the $d$-quark in nuclear matter there are two more quarks, namely $u$-quark and $d$-quark. In fact, we are talking about the implementation of a four-fermion interaction initiated by a vector boson. This is essential for initiating the "relaxation" quark-cumulative dynamics of nuclear matter, in which, as indicated above, the principle of least action is realized in the formation of the final decay products. Therefore, using the constant $G_F$ is quite justified in this case. We also believe that in this case, the manifestation of possible anomalies in the kinematics of cumulative processes is facilitated by a decrease in the stability of the surface of the nucleus due to a violation of the nucleonic structure of nuclear matter [20].

In addition to the $e^-$-catalytic initiation of radioactive decay processes, which we will define as the "1st type $e^-$-catalysis," the nuclear chemical processes under consideration can be initiated in a low-temperature non-equilibrium plasma through direct interaction of nuclei located in the plasma with neutral nuclei $^1n_{isu}$ and $^2n_{isu}$ formed in the plasma, whose characteristic lifetimes are quite long (see above). We will define these processes as "2nd type -catalytic processes."

INITIATION OF NUCLEI TRANSMUTATION AND ARTIFICIAL RADIOACTIVITY PROCESSES: 2nd TYPE -CATALYSIS

According to the experimental findings [16] and [17], the nuclei $^1n_{isu}$ and $^2n_{isu}$ upon merging with the original target nuclei and forming "composite" nuclei not only imparted an energy of ~10 MeV into the nuclear matter of the original nuclei but could also lead to the formation of "composite" nuclei as $isu$-nuclei in unbalanced $isu$-state, resulting in instability. These unstable "composite" nuclei, formed under the influence of deuterium-containing glow discharge plasma on impurity Pt nuclei in the Pd cathode, could decay, as indicated by processes (18)-(20) reported [16], resulting in the formation of heavy isotopes elements (particularly tungsten isotopes) and the emission of N-14 or C-12 as "light" nuclei:

$$^{195}_{78}\text{Pt} + ^2\text{n}_{\text{isu}} \rightarrow ^{197}_{78}\text{Pt}^*_{\text{isu}} \rightarrow ^{14}_{7}\text{N} + ^{183}_{74}\text{W} + 3e^- + 3\tilde{\nu} + \nu\tilde{\nu} + Q(23.83\,\text{MeV}), \tag{11}$$

$$^{198}_{78}\text{Pt} + ^2\text{n}_{\text{isu}} \rightarrow ^{200}_{78}\text{Pt}^*_{\text{isu}} \rightarrow ^{14}_{7}\text{N} + ^{186}_{74}\text{W} + 3e^- + 3\tilde{\nu} + \nu\tilde{\nu} + Q(22.96\,\text{MeV}), \tag{12}$$

$$^{A}_{78}\text{Pt} + ^2\text{n}_{\text{isu}} \rightarrow ^{A+2}_{78}\text{Pt}^*_{\text{isu}} \rightarrow ^{12}_{6}\text{C} + ^{A-10}_{74}\text{W} + 2e^- + 2\tilde{\nu} + \nu\tilde{\nu} + Q_A, \tag{13}$$

where $A$ = 190, 192, 194 and 196, and the energy release values for the indicated $A$ during the formation of tungsten isotopes 180, 182, 184 and 186 are equal to 25.45, 25.09, 24.06, and 22.98 MeV, respectively. Nuclear reactions of this type involving initially nonradioactive isotopes, when during the decay of a compound nucleus the indicated "heavy" light nuclei are emitted in addition to the heavy isotope, are not characteristic of the processes of initiation of artificial radioactivity when the target is irradiated with protons and α-particles. Within the framework of modern ideas of this type, the products of nuclear collisions should rather be realized in r- processes of nucleosynthesis during supernova explosions and during the merger of neutron stars [25–27].

The so-called cluster radioactivity, which is the phenomenon of spontaneous emission of nuclear fragments (clusters) more massive than α-particle, was first discovered in 1984 [28, 29]. When studying the α-activity of nuclei $^{223}_{88}\text{Ra}$, the authors [28] discovered that these nuclei, instead of α-particles, sometimes (with a probability lower by almost 10 orders of magnitude) emit nuclei $^{14}_{6}\text{C}$; therefore, a "cluster" decay is realized:

$$^{223}_{88}\text{Ra} \rightarrow ^{14}_{6}\text{C} + ^{209}_{82}\text{Pb} + Q\,(31.84\,\text{MeV}). \tag{14}$$

Currently, more than 20 nuclei from $^{114}_{56}\text{Ba}$ to $^{242}_{96}\text{Cm}$, have been observed to occasionally emit nuclei more massive than an $α$ particle during the process of radioactive decay. Among the emitted nuclei $^{14}_{6}\text{C}$, $^{20}_{8}\text{O}$, $^{24}_{10}\text{Ne}$, $^{26}_{10}\text{Ne}$, $^{28}_{12}\text{Mg}$, $^{30}_{12}\text{Mg}$, $^{32}_{14}\text{Si}$, and $^{34}_{14}\text{Si}$ were recorded [30]. However, the probability of such processes is extremely low, being 10–17 orders of magnitude less than the probability of emitting $α$ particles by the same nuclei.

Therefore, the unusual nature of the phenomenon of initiation of nuclear chemical processes under conditions of low-temperature nonequilibrium plasma, and even when emitting "light" nuclei C-12, N-14, O-16 in addition to heavy isotopes, is clearly associated with the formation of a disrupted *isu*-structure of nuclear matter of compound nuclei in the nuclear reactions considered [16, 17], followed by the implementation of a kind of "quark-cumulative mechanisms" of the transformations taking place.

Returning to the analysis of the nuclear chemical processes considered [16, 17], initiated under conditions of a glow discharge by the action of plasma flows on samples of Pd, Ni [16] and Pb [17], containing impurity elements, a remarkable "consistency" is observed in the ratios of isotopes of elements formed within the volume of samples of elements. This consistency is evident when estimating the relative content of isotopes of a particular element on the number of pulses recorded during ICP MS analysis. The notable example is the formation of W isotopes in Pd and Pb samples, the content of which in the initial Pd sample was generally recorded at the background level. At the same time, the close correspondence of the relative proportions of the resulting W isotopes to the known natural ratios of this element, and not to the proportions of the corresponding Pt and Pb isotopes, which, according to the nuclear chemical processes presented [16, 17], were considered as basic parent nuclei, was unexpected. This circumstance, along with the obvious differences in the values of phase volumes (due to differences in the total number of final particles formed in accordance with the stoichiometry in specific nuclear reactions) for the W isotopes under consideration, may also indicate that nuclear chemical processes go beyond the kinematic region prohibited by the laws of energy-momentum conservation for free "particle-particle" collisions. Thus, the specific nature of transformations in such processes, which can be characterized as quark-cumulative, is highlighted.

It is important to consider that the dynamics of relaxation rearrangement of the broken nucleon *isu*-structure of the compound nucleus can significantly influence the formation of daughter isotopes in the nuclear chemical processes studied in [13] and [16]. The rearrangement of the *isu*-structure of a compound nucleus is accompanied not by processes of emission of gamma quanta, as is the case with the proton-neutron organization of nuclear matter, but under the conditions of the decisive role of weak nuclear interactions by processes of emission of neutrino-anitrinoneutrino pairs or Gamow-Schönberg URCA processes [31]. This aspect highlights the safety of low energy nuclear chemical processes for the environment. Moreover, it underscores the reason for the existence of life on our planet, as these low-energy synthesis and transmutation processes of chemical elements are inherent in the life of any organism [22, 32].

CONCLUSIONS

The introduction of the concept of initiated processes of the $\beta$-decay and electron capture raises a number of controversial issues. One of these concerns a seemingly resolved problem—the cause of the deficit in solar neutrinos. In Sun's thermonuclear reactions, when a helium atom is formed during combustion from four hydrogen atoms with the release of 26.7 MeV of energy, two neutrinos are produced, carrying away approximately 0.6 MeV of energy each (in this case, the energy spectrum of solar neutrinos was found to be quite extended, up to energies of 14 MeV) [33]. Based on these data and the known value of the Sun's luminosity, the flux of solar neutrinos incident on the Earth was estimated to be about $10^{11}$ particles/(cm$^2$ s).

To register such flows, they usually use neutrino detectors, such as chlorine-argon and gallium-germanium ones. It is believed that in such detectors solar neutrinos $\nu_{he}$ with quite high energy $E_{vhe}$ interact with basic isotopes. Fluxes $\nu_{he}$ in chlorine-argon detectors at $E_{vhe} > 0.814$ MeV interact with isotopes $^{37}_{17}Cl$, which turn into isotopes $^{37}_{18}Ar$ when emitting an electron $e^-$ and antineutrino $\tilde{\nu}$. In gallium-germanium detectors, fluxes $\nu_{he}$ at $E_{vhe} > 0.232$ MeV interact with isotopes $^{71}_{31}Ga$, which turn into isotopes $^{71}_{32}Ge$ with emission $e^-$ and $\tilde{\nu}$. The neutrino flux is measured in solar neutrino units (SNU): this unit corresponds to the neutrino flux at which $10^{-36}$ reactions per second per selected atom. For the gallium-germanium experiment, the observed neutrino flux was about 70 SNU, while theory predicted a value of 122 SNU. For the chlorine-argon experiment, the observed value was about 2.5 SNU, that is, only about a third of the theoretical value of 8.0 SNU.

If we keep in mind the results of the present work, the shortage of the so-called "neutrino fluxes" "measured" by the methods used should be recorded. Really, in accordance with the ideas developed in this work, the nuclear transformations under consideration are possible only through the interaction of neutrino fluxes of the indicated energies with previously activated in processes (1) nuclei $^{37}_{17}Cl$ and $^{71}_{31}Ga$, i.e. specifically, with nuclei $^{37}_{16}S_{isu}$ and $^{71}_{30}Zn_{isu}$:

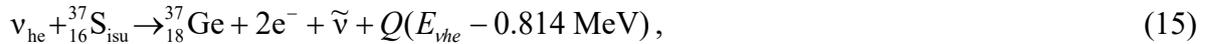
$$\nu_{he} + ^{37}_{16}S_{isu} \rightarrow ^{37}_{18}Ge + 2e^- + \tilde{\nu} + Q(E_{vhe} - 0.814 \text{ MeV}), \qquad (15)$$

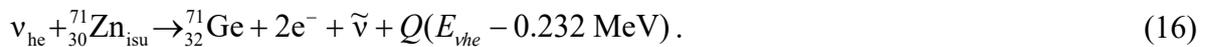
$$\nu_{he} + ^{71}_{30}Zn_{isu} \rightarrow ^{71}_{32}Ge + 2e^- + \tilde{\nu} + Q(E_{vhe} - 0.232 \text{ MeV}). \qquad (16)$$

This raises a natural question: what is the proportion of such activated *isu*-nuclei are located, respectively, in arrays of tetrachlorethylene and gallium used in chlorine-argon and gallium-germanium detectors, respectively. Obviously, this fraction depends on the specific radiation background at the location of the detector and can be purposefully increased. It is possible that when solving a set of emerging issues, it will be necessary to revise the used values for the cross section for nuclear processes initiated by neutrinos. We also point out that the recent problems with the uncertainty of the cross section for neutrino absorption by gallium in installations for detecting solar neutrinos [34] may be relevant to the issue under discussion. We believe that a more comprehensive discussion of the range of issues raised can provide greater physical clarity to the current resolution of the solar neutrino deficit problem. This includes considering the contribution to the total neutrino flux from neutrinos of two other types (muon and tau neutrinos), which are not observable in neutrino detectors designed for electron neutrinos, despite the fact that all three types of neutrinos can undergo transformation into each other (known as "neutrino oscillations").